# Data Hiding using Graphical Code based Steganography Technique

Debajit Sensarma[*1], Samar Sen Sarma[*2]

[#1]*Research Fellow, Department of Computer Science & Engineering, University of Calcutta*
Kolkata, INDIA

[#2]*Professor, Department of Computer Science & Engineering, University of Calcutta*
Kolkata, INDIA

**Abstract**— *Data hiding has received much attention due to rapid development of internet and multimedia technologies where security of information is a very important concern. This is achieved by Steganography, which is the art or science of hiding data into another data, so that human eyes cannot catch the hidden information easily. There are many ways to hide information-like inside an image, text, audio/ video etc. Among them image steganography is a very attractive research area. The goal is to transmit a data within a modified image (called stego-image) by minimizing the number of bit flips. In this paper, a new steganography technique has been proposed using Graphical codes and also comparison with steganography technique using BCH codes has been studied.*

## I. INTRODUCTION

The word "Steganography" comes from the Greek words "Steganos" meaning covered, concealed, or protected and "Graphein" meaning writing [1]. So, this is an art of concealing a file, message, image or video within another file. The steganography techniques, based on cover medium, are mainly of four categories- i) image steganography, ii) steganography in audio, iii) video steganography and iv) text steganography [2]. Image steganography is post popular among them. Besides this there are three types of steganography techniques, they are- i) pure steganography, ii) secret key steganography, iii) public key steganography [3].

It is well known from [9], [12] that the set of all even sub-graphs of connected graph G of n vertices with e edges forms a binary linear code C, with parameters [n, e-n+1, g], where g is the girth of the graph G. In [9, 10, 11], the concepts of graph theoretic codes like- cut-set codes, circuit codes, augmented circuit codes are given and the decoding procedure (i.e. majority logic decoding) along with the efficiency of this graph theoretic block codes has been described. But it is not clear if the augmented codes are graphical or not. Later, in [12, 13] authors noticed this issue and proposed a new method to generate augmented graphical code and their decoding using combinatorial optimization, which is more efficient than majority decoding procedure.

In this paper, we mainly considered image steganography technique. There are various types of embedding techniques like- LSB steganography, pixel value differencing steganography, RGB based steganography, mapping based steganography, code based steganography etc. [2]. We have proposed a Graphical code based image steganography technique. Also, compared the performance of this method with steganography technique using BCH code and it can be seen that in some cases the embedding rate and embedding efficiency of the technique based on Graphical Code is better than the technique based on BCH code.

The paper is organized as follows: In section II basics of coding is given. Section III describes the application of error correcting codes in steganography. Some related works are given in section IV. Section V describes the Graphical Codes in brief. Proposed method of steganography is given in section VI. Section VII contains an illustrative example. Some results related to the proposed method are given in section VIII and section IX concludes the paper by giving some future scope.

## II. BASICS OF CODING

In this section, a brief overview of some concepts regarding coding is given. A code S is any finite set, which is he subset of the space of all n-bit vectors of the form $c = \{c_1, c_2,..., c_n\} \in \{0, 1\}^n$. The vectors in S are called codewords. As in the most of the applications binary codes are used, we restrict ourselves to this case. The set $\{0, 1\}^n$ forms a linear vector space over the filed GF(2) [5, 6] denoted as $F_2$.

Hamming distance d(c, y) between the bit strings $c = c_1, c_2,..., c_n$ and $y = y_1, y_2,..., y_n$ is the number of positions in which theses strings differ, that is the number of j , (j= 1,2, ..., n) where $c_j \neq y_j$ [7]. For codeword c and $S \subset F_2^n$ $d(c, S) = \min_{x \in S} (c, x)$. The covering radius $\rho$ of S is defined as $\rho = \max_{c \in F_2^n} d(c, S)$.

A linear code is an error correcting code for which any linear combination of codewords is also a codeword. It is mainly of two categories- block code and convolutional code. Linear block code of length n and k information digits (also called rank) is a linear subspace S with dimension k of vector space $F_q^n$ where $F_q$ is a finite field with q elements. If q=2, then S is referred to as binary (n, k) code. There are $2^k$ distinct messages. The encoder transforms each input message to a binary n tuple c, with n > k according to certain rule. So, for $2^k$ messages there are $2^k$ codewords. The set of $2^k$ codewords is called block code. A (n, k) linear code S is a k-dimensional subspace of the vector space $S^n$ of all binary n-tuples. It is possible to find k-linearly independent codewords from them and it can be written in k X n rows form of a matrix called Generator matrix. For any k X n matrix G with k linearly independent rows, there exists a (n-k) X n matrix H with n-k linearly independent rows such that any vector in the row space of G is orthogonal to the rows of H and any vector that is orthogonal to the rows of H is the row space of G. So, an n-tuple x is a codeword if and only if $x.H^T = 0$ ($H^T$ = Transpose of matrix H). The matrix H is called a Parity Check matrix. The $2^{n-k}$ linear combinations of the rows of matrix H form a (n, n-k) linear code [6].

For any $c \in F_2^n$, the vector $s = c.H^T \in F_2^{n-k}$ is called syndrome of the codeword c. For each syndrome $s \in F_2^{n-k}$, the set C(s) = {$c \in F_2^n$ | c.$H^T$ =s} is called a coset. Every coset can be written as C(s) = c + S, where $c \in C(s)$ arbitrarily. Thus there are $2^{n-k}$ distinct cosets, each consisting of $2^k$ vectors. The member of the coset with minimum Hamming distance is called coset leader ($l_s$), although it is not necessarily unique.

### III. ERROR CORRECTING CODES IN STEGANOGRAPHY

One important kind of steganography protocol can be defined from coding theory point of view. Though, Error Correcting codes are mainly used for detecting and correcting error or erasures. In [15] a relation between perfect code and maximum length embeddable code are shown. Most of the codes used in steganography are linear in nature and good steganography protocols are designed based Parity Check matrix.

Let, n and p be the positive integers, $p \leq n$ and S is a finite set. An embedding or retrieval steganography protocol [n, p] over S, maps m: $S^p \times S^n \rightarrow S^n$ and u: $S^n \rightarrow S^p$ such that u (m(s, c)) = s for all $s \in S^p$ and $c \in S^n$. The maps m and u are the embedding and retrieval maps respectively. Here $\rho$ = max {d (c, m(s, c)) | $s \in S^p$ and $c \in S^n$}.

The embedding map of [n, p] embedding, retrieval steganography protocol with radius $\rho$ is called [n, p, $\rho$] protocol which allows to hide p data symbols into a string of n cover symbols by changing maximum of $\rho$ cover symbols [16]. The syndrome map u: $S^n \rightarrow S^p$ is called retrieval map of a [n, p, $\rho$] linear steganography protocol.

The goal of image steganography is to least modify the image and embed data in the image such that the syndrome calculated at receiver side is the required hidden data. Let, $c \in F_2^n$ is the vector extracted from the cover medium or the image and $m \in F_2^{n-k}$ is the data to be hidden. The data is inserted into c by least modification possible. So, modify c to $v \in F_2^n$ such that $v.H^T = m$, where H is the Parity Check matrix. For constructing v, vector w (with minimum weight) is needed whose syndrome is m-c.$H^T$ and v = w+ c. It leads to v.$H^T$=c. $H^T$ + w. $H^T$ = c.$H^T$ + m- c.$H^T$ =m.

There are two parameters helps in performance evaluation of [n, p, $\rho$] protocol. Firstly, Embedding Rate (ER) = p/n and Embedding Efficiency (EF) = p/$\rho$, i.e. the number of random data bits embedded per embedding change.

### IV. RELATED WORKS

In this section some related works on some steganography techniques based on codes are described in brief. In [17] author proposed a steganography technique $F_5$, based on matrix encoding idea which has improved embedding efficiency [18]. Here m bit data can be embedded in $2^m-1$ cover symbols by changing at most one symbol. The concept of Hamming Code is used in this scheme. Authors in [19] take the advantages of Reed Solomon Codes (RS codes) which according to [20] is a good tool for steganography technique. They have proposed a matrix embedding technique based on RS codes which allows easy way to solve the bounded syndrome problem. Next, in [21] authors used wet paper codes using random linear codes of small co-dimension which has much better embedding efficiency. In [16] authors proposed a steganography technique using BCH [$2^m-1$, $2^m-2m-1$] code {m=3,...7} and majority logic decoding. Next, the authors of [15] proposed a new technique called product perfect code and its application to steganography. In [22], a robust (non-fragile) steganography technique has been introduced based on matrix encoding using self-synchronizing variable length T-codes and RS codes for obtaining compressed image from the original image and robustness against transmission error. Besides this, more steganography techniques has been listed in [23-25].

### V. GRAPHICAL CODES

In this section the concepts of Graphical Codes are given in a nutshell. For general graph theory background reader can refer to [8]. Let, G (V, E) be a connected undirected graph with V = {$v_1$,...$v_n$} vertices and E = {$e_1$, ...$e_m$} edges. An Euler sub-graph of a graph G is a sub-graph g, in which every vertex has even degree. According to [9] Euler sub-graph is either a circuit or edge disjoint union of circuits. Every sub-graph g can be described using a binary characteristic vector g = ($g_1$,..., $g_m$), where $g_i$ = 1 if $e_i$ is an edge of g and $g_i$ = 0 otherwise ($1 \leq i \leq m$). There are two subspaces associated with every graph. Circuit Space ($W_\Gamma$) or Cycle Space generated by all circuits or edge disjoint union of circuits of G and Cut-set Space or bond space ($W_S$) generated by every cut-sets or edge disjoint union of cut-sets. Let, t is a spanning tree of G. So, each edge not in t forms circuit with t and the characteristic vectors of these (m-n+1) circuits are linearly independent and this linearly independent row vectors forms a matrix of dimension $W_\Gamma$, is called Fundamental Circuit matrix. Similarly, each edge of t is associated with the cut-set of G and has (n-1) linearly independent cut-sets which also generates a matrix of dimension $W_S$, called Fundamental Cut-set matrix. It can be shown from [8] that, $W_\Gamma$ and $W_S$ are orthogonal to each other. So, from the above description, it is clear that the Circuit Space (also Cut-set Space) of G forms binary linear code C, with parameters [n, n-m+1, g] (also [n, n-1, g]), where g is the girth of G. The code C is termed as Graphical Code. Basically, in this paper Circuit code is considered and here Fundamental Circuit matrix acts as generator matrix and Fundamental Cut-set matrix acts as Parity Check matrix. In [12, 13], the circuit code is denoted as $C_E$ (G). These codes are firstly studied in [9, 10, 11]. The objective was to show that possibility of augmenting the Graphical Code (also called even graphical code [13]) to the larger dimension by keeping minimum distance unchanged and also to provide a decoding algorithm for these codes. Next, in [12, 13], authors shows an improvement over the decoding procedure using combinatorial optimization technique. T-join of the graph G, where $T \subseteq V$, is a subset of edges E, which has odd degree at every vertex in T and even degree in every other vertex. Necessarily T has even cardinality. The smallest possible cardinality of T-join will be denoted by $\tau$ (G, T). According to [14] the covering radius is equals the maximum vertex join number $\tau$ (G), where $\tau$ (G) = $\max_T$ ($\tau$ (G, T)), which is the largest size of the minimum T-join for any even vertex set T. For further information on covering radius of Graphical Code, reader can refer [14].

### VI. PROPOSED STEGANOGRAPHY TECHNIQUE

Let, $t(X) \in F_2^n$, be the polynomial representation of the extracted vector from the cover image. $V(X) \in F_2^n$, be the polynomial representation of vector of stego-medium and $m(X) \in F_2^{n-k}$, is the polynomial representation of vector of information to be embedded. The aim is to change $t(X)$ to $V(X)$ such that $m(X)$ is embedded in $V(X)$ and least number of bit positions are flipped.

Suppose, $e(X)$ be the flip pattern representing the number of bit positions flipped. So, the stego-data $V(x) = t(X) + e(X)$ ... (1).

From the relationship between m, v it can be seen that, $v.H^T = m$ ... (2).

So, from eq. (1) and eq. (2) we can get,
$v.H^T = m$
or, $(t + e).H^T = m$

or, $t.H^T + e.H^T = m$
or, $e.H^T = m - t.H^T$

So, a vector e has to be found whose syndrome is $s = m - t.H^T$ ... (3).

**Algorithm 1:** Embedding of Data

**Inputs:** Data m, image block t.
**Output:** Stego-image

**Step 1:** Compute s from eq. 3.
**Step 2:** If s = 0, then data is already hidden and stop, else goto step 3.
**Step 3:** Find vector e from leader-syndrome table [26] which is also the minimum T-join corresponding to the syndrome s.
**Step 4:** Modify the cover medium t, with t + e and generate vector v.
**Step 5:** If no more data to be embedded then stop, else goto step 1.

**Algorithm 2:** Extraction of Data
**Input:** Stego-image
**Output:** Hidden data

**Step 1:** Read n symbols v from stego-image.
**Step 2:** Calculate data m from eq. 2.
**Step 3:** If there are no symbols left then stop, else goto step 1.

### VII. ILLUSTRATION WITH AN EXAMPLE

Complete graph $K_n$ gives rise to a binary linear code with parameters [$^nC_2$, (n-1)(n-2)/2, 3] where number of edges = $^nC_2$, number of vertices = n and girth = 3 [12]. We have constructed the example with n=5. For n=5, it is [10, 6, 3] code with covering radius = 2 [14].

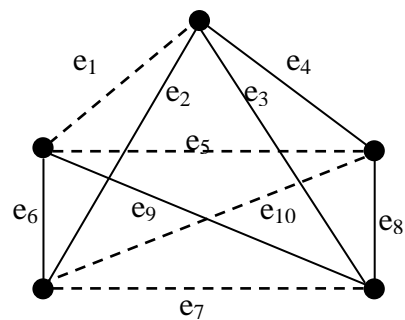

Figure 1: 5-vertex Complete Graph

$$\begin{array}{c|cccccccccc} & e_2 & e_3 & e_4 & e_5 & e_6 & e_7 & e_8 & e_9 & e_{10} \\ e_1 & 1 & 0 & 0 & 1 & 0 & 0 & 0 & 0 & 1 \\ 1 & 0 & 1 & 0 & 1 & 0 & 1 & 0 & 0 & 1 \\ 1 & 0 & 0 & 1 & 1 & 0 & 0 & 0 & 0 & 0 \\ 0 & 0 & 0 & 0 & 1 & 1 & 0 & 0 & 0 & 1 \\ 0 & 0 & 0 & 0 & 0 & 0 & 1 & 1 & 0 & 1 \\ 0 & 0 & 0 & 0 & 1 & 0 & 1 & 0 & 1 & 1 \end{array}$$

Figure 2: Fundamental Circuit Matrix of the graph in Figure 1

$$\begin{pmatrix} e_1 & e_2 & e_3 & e_4 & e_5 & e_6 & e_7 & e_8 & e_9 & e_{10} \\ 1 & 1 & 1 & 1 & 0 & 0 & 0 & 0 & 0 & 0 \\ 0 & 1 & 1 & 1 & 1 & 1 & 0 & 0 & 1 & 0 \\ 0 & 1 & 1 & 0 & 0 & 1 & 0 & 1 & 1 & 1 \\ 0 & 0 & 1 & 0 & 0 & 0 & 1 & 1 & 1 & 0 \end{pmatrix}$$

Figure 3: Fundamental Cut-Set Matrix of the graph in Figure 1

**TABLE I: Syndrome-Coset Leader Table**

| Syndrome | Coset Leader |
|---|---|
| 0001 | $e_7$ |
| 0010 | $e_{10}$ |
| 0011 | $e_8$ |
| 0100 | $e_5$ |
| 0101 | $e_5 + e_7, e_6 + e_8, ...$ |
| 0110 | $e_6$ |
| 0111 | $e_9$ |
| 1000 | $e_1$ |
| 1001 | $e_1 + e_7, e_3 + e_6, ...$ |
| 1010 | $e_1 + e_{10}, e_2 + e_5, ...$ |
| 1011 | $e_1 + e_8, e_4 + e_9, ...$ |
| 1100 | $e_4$ |
| 1101 | $e_4 + e_7, e_3 + e_{10}, ...$ |
| 1110 | $e_2$ |
| 1111 | $e_3$ |

**Ex 1:** Suppose we want to hide the 4 bit data (1100) into the vector (1101111011) using minimal bit flip.

**Sol:**
From given vector the codeword of the graphical code is:

$e_1\ e_2\ e_3\ e_4\ e_5\ e_6\ e_7\ e_8\ e_9\ e_{10}$
1   1   0   1   1   1   1   0   1   1

From Algorithm 1:

$s_1 = e_2 + e_3 + e_4 + e_1 = 1 + 0 + 0 + 1 = \mathbf{1}$
$s_2 = e_2 + e_3 + e_4 + e_6 + e_9 + e_5 = 1 + 0 + 0 + 1 + 1 + 1 = \mathbf{1}$
$s_3 = e_2 + e_3 + e_6 + e_8 + e_9 + e_{10} = 1 + 0 + 1 + 0 + 1 + 1 = \mathbf{0}$
$s_4 = e_3 + e_8 + e_9 + e_7 = 0 + 0 + 1 + 1 = \mathbf{0}$

From eq. (3)

$s = (1100-1100) = (0000)$

So, the data is already hidden in the stego-image.

**Ex 1:** Suppose we want to hide the 4 bit data (1010) into the vector (1101111011) using minimal bit flip.

**Sol:**
From given vector the codeword of the graphical code is:

$e_1\ e_2\ e_3\ e_4\ e_5\ e_6\ e_7\ e_8\ e_9\ e_{10}$
1   1   0   1   1   1   1   0   1   1

From Algorithm 1:

$s_1 = e_2 + e_3 + e_4 + e_1 = 1 + 0 + 0 + 1 = \mathbf{1}$
$s_2 = e_2 + e_3 + e_4 + e_6 + e_9 + e_5 = 1 + 0 + 0 + 1 + 1 + 1 = \mathbf{1}$
$s_3 = e_2 + e_3 + e_6 + e_8 + e_9 + e_{10} = 1 + 0 + 1 + 0 + 1 + 1 = \mathbf{0}$
$s_4 = e_3 + e_8 + e_9 + e_7 = 0 + 0 + 1 + 1 = \mathbf{0}$

From eq. (3)

$s = (1010-1100) = (0110)$
From Table I:

Vector e = (0000010000)

So, modified vector v= (1101111011+ 0000010000) = (1101101011).

## VIII. RESULTS

Here, the embedding rate (ER = k/n) and embedding efficiency (EF = k/$\rho$) of the steganography technique based on BCH codes, Circuit codes and augmented circuit codes are given. Here ($K_B$, $K_C$, $K_A$), ($\rho_B$, $\rho_C$, $\rho_A$), ($ER_B$, $ER_C$, $ER_A$) and ($EF_B$, $EF_C$, $EF_A$) are the hidden data can be embedded, covering radius, embedding rate and embedding efficiency corresponding to BCH codes, graphical codes and augmented graphical codes respectively, 'd' is the Hamming distance between two codewords. The data are taken from [9] and the comparison is shown in table II.

## IX. CONCLUSIONS

In this paper, a new Graphical Code based steganography technique has been proposed. Here, in Algorithm 1, data is embedded with minimum number of bit flip and Algorithm 2 gives the procedure of hidden data extraction efficiently. From table II, it can be seen that, in some cases using Graphical Codes gives better embedding efficiency than using BCH codes. The performance is measured by varying the minimum distance between two codewords and covering radius of the codes.

In future we will try to compare the proposed method with other steganography techniques and will try to improve the security of the technique by using cryptographic secret key or public key such that the proposed method can be applied to various real world problems. Furthermore, we will try to design a unified steganography technique which will be applicable to image, audio, video and text steganography techniques etc.

## ACKNOWLEDGMENT

The authors would like to thank University Of Calcutta, West Bengal, India, Department of Science & Technology (DST), New Delhi, for financial support and the reviewers for their constructive and helpful comments and specially the Computer without which no work was possible.

TABLE II: Comparison between Embedding rate and Embedding Efficiency

| n | $K_B$ | $K_C$ | $K_A$ | d | $\rho_B$ | $\rho_C$ | $\rho_A$ | $ER_B$ | $ER_C$ | $ER_A$ | $EF_B$ | $EF_C$ | $EF_A$ |
|---|---|---|---|---|---|---|---|---|---|---|---|---|---|
| 15 | 4 | 5 | 4 | 3 | 1 | 2 | 1 | 0.267 | 0.33 | 0.267 | 4 | 2.5 | 4 |
|  | 8 | 9 | 9 | 5 | 3 | 2 | 2 | 0.53 | 0.6 | 0.6 | 2.67 | 4.5 | 4.5 |
|  | 10 | 12 | 11 | 7 | 5 | 6 | 5 | 0.67 | 0.8 | 0.733 | 2 | 2 | 2.2 |
| 21 | 9 | 13 | 11 | 5 | 3 | 5 | 4 | 0.43 | 0.62 | 0.52 | 3 | 2.6 | 2.5 |
|  | 15 | 16 | 14 | 7 | 6 | 8 | 5-6 | 0.71 | 0.76 | 0.67 | 2.5 | 2 | 2.8-2.33 |
| 31 | 5 | 8 | 7 | 3 | 1 | 2 | 2 | 0.16 | 0.29 | 0.23 | 5 | 4 | 3.5 |
|  | 10 | 17 | 14 | 5 | 3 | 5-6 | 4 | 0.32 | 0.55 | 0.45 | 3.33 | 3.4-2.83 | 3.5 |
|  | 15 | 21 | 19 | 7 | 4-5 | 7-9 | 6-7 | 0.48 | 0.68 | 0.61 | 3.75-3 | 3-2.33 | 3.17-2.71 |
| 35 | 24 | 18 | 15 | 5 | 8-9 | 5-6 | 4-5 | 0.69 | 0.51 | 0.43 | 3-2.67 | 3.6-3 | 3.75-3 |
|  | 27 | 23 | 21 | 7 | 10-12 | 7-9 | 7-8 | 0.77 | 0.66 | 0.6 | 2.7-2.25 | 3.29-2.56 | 3-2.63 |
| 45 | 16 | 22 | 17 | 5 | 4-5 | 6 | 4-5 | 0.36 | 0.49 | 0.38 | 4-3.2 | 3.67 | 4.25-3.4 |
|  | 22 | 29 | 24 | 7 | 6 | 9-11 | 7-8 | 0.49 | 0.64 | 0.53 | 3.67 | 3.22-2.64 | 3.43-3 |
| 55 | 40 | 25 | 19 | 5 | 13-17 | 7 | 5 | 0.72 | 0.45 | 0.35 | 3.08-2.35 | 3.57 | 3.8 |
| 63 | 6 | 11 | 8 | 3 | 1 | 3 | 2 | 0.09 | 0.18 | 0.13 | 6 | 3.67 | 4 |
|  | 12 | 28 | 21 | 5 | 3 | 7-8 | 5-6 | 0.19 | 0.44 | 0.33 | 4 | 4-3.5 | 4.2-3.5 |
|  | 18 | 38 | 30 | 7 | 4-5 | 11-13 | 8-9 | 0.29 | 0.6 | 0.48 | 4.5-3.6 | 3.45-2.92 | 3.75-3.33 |